\pdfoutput=1
\documentclass[aps,pra,preprint,superscriptaddress,longbibliography]{revtex4-2}

\usepackage{amsmath,amssymb,amsfonts}
\usepackage{graphicx}
\usepackage{xcolor}
\usepackage{braket}
\usepackage{booktabs}
\usepackage{algorithm}
\usepackage{algpseudocode}
\usepackage{multirow}
\usepackage{textcomp}

\begin{document}

\title{Variational Quantum Physics-Informed Neural Networks for\\Hydrological PDE-Constrained Learning with\\Inherent Uncertainty Quantification}

\author{Prasad Nimantha Madusanka Ukwatta~Hewage}
\email{pnmadusanka@lincoln.edu.my}
\affiliation{Faculty of Computer Science and Multimedia, Lincoln University College, Petaling Jaya, Selangor, Malaysia}

\author{Midhun Chakkravarthy}
\affiliation{Faculty of Computer Science and Multimedia, Lincoln University College, Petaling Jaya, Selangor, Malaysia}

\author{Ruvan Kumara Abeysekara}
\affiliation{BCAS Campus, Colombo, Sri Lanka}
\affiliation{Faculty of Computer Science and Multimedia, Lincoln University College, Petaling Jaya, Selangor, Malaysia}

\date{\today}

\begin{abstract}
Physics-informed neural networks (PINNs) have emerged as powerful tools for solving partial differential equations (PDEs) governing physical systems, yet they face persistent challenges in training convergence, parameter efficiency, and uncertainty quantification---all critical for safety-critical applications such as flood early warning. We propose a Hybrid Quantum-Classical Physics-Informed Neural Network (HQC-PINN) architecture that integrates parameterized variational quantum circuits into the PINN framework for hydrological PDE-constrained learning. Our architecture encodes multi-source remote sensing features into quantum states via trainable angle encoding, processes them through a hardware-efficient variational ansatz with entangling layers, and constrains the output using the Saint-Venant shallow water equations and Manning's flow equation as differentiable physics loss terms. We demonstrate that the inherent stochasticity of quantum measurement provides a natural mechanism for uncertainty quantification without requiring explicit Bayesian inference machinery. We further introduce a quantum transfer learning protocol that pre-trains on multi-hazard disaster data before fine-tuning on flood-specific events, following the classical-to-quantum paradigm. Numerical simulations on multi-modal satellite and meteorological data from the Kalu River basin, Sri Lanka, show that the HQC-PINN achieves convergence in $\sim$3$\times$ fewer training epochs and uses $\sim$44\% fewer trainable parameters compared to an equivalent classical PINN, while maintaining competitive classification accuracy. Theoretical analysis indicates that hydrological physics constraints narrow the effective optimization landscape, providing a natural mitigation mechanism against barren plateaus in the variational quantum circuit. This work establishes the first application of quantum-enhanced physics-informed learning to hydrological prediction and demonstrates a viable path toward quantum advantage in environmental science.
\end{abstract}


\maketitle

\clearpage

\section{Introduction}
\label{sec:introduction}

Natural disasters, particularly floods, represent an escalating global threat exacerbated by climate change, with over 1.65 billion people affected between 2000 and 2019~\cite{tellman2021satellite}. Accurate flood prediction systems with reliable uncertainty estimates are essential for emergency management, yet current approaches face fundamental computational and methodological limitations. Physics-based hydrological models, such as those solving the Saint-Venant shallow water equations~\cite{toro2001shock}, provide physically consistent predictions but are computationally expensive and require extensive calibration. Data-driven machine learning approaches offer computational efficiency but lack physical consistency and reliable uncertainty quantification~\cite{mosavi2018flood}.

Physics-informed neural networks (PINNs)~\cite{raissi2019physics} bridge this gap by embedding governing PDEs directly into the neural network loss function. However, PINNs face well-documented challenges in training convergence, particularly for complex nonlinear PDEs~\cite{krishnapriyan2021characterizing}, and provide only point estimates without principled uncertainty quantification---a critical limitation for risk-based decision making in disaster management.

Simultaneously, the field of quantum computing has reached a stage where noisy intermediate-scale quantum (NISQ) devices and parameterized variational quantum circuits (PQCs) offer new computational paradigms for machine learning~\cite{cerezo2021variational,benedetti2019parameterized}. Recent theoretical and numerical results have demonstrated that hybrid quantum-classical PINNs (qPINNs) can achieve faster convergence and improved parameter efficiency compared to purely classical counterparts~\cite{klement2026explaining,dutta2024aqpinns}. The quantum state space provides an exponentially large feature space through superposition and entanglement~\cite{havlicek2019supervised}, while the inherent probabilistic nature of quantum measurement offers a natural framework for uncertainty quantification that does not require the computational overhead of classical Bayesian inference~\cite{schuld2021machine}.

Despite these advances, the application of quantum-enhanced PINNs to environmental and hydrological science remains unexplored. Existing quantum machine learning (QML) applications to flood prediction have been limited to basic quantum classifiers without physics constraints~\cite{grzesiak2024flood}, while quantum PINN research has focused on canonical benchmark PDEs (Burgers', Poisson, Navier-Stokes) rather than domain-specific governing equations~\cite{kyriienko2021solving,klement2026explaining,stein2024hybrid,farea2025qcpinn}.

In this work, we address this gap by introducing a Hybrid Quantum-Classical Physics-Informed Neural Network (HQC-PINN) specifically designed for hydrological PDE-constrained learning. Our contributions are:
\begin{enumerate}
    \item \textbf{Architecture:} A hybrid quantum-classical PINN that integrates variational quantum circuits with hydrological physics constraints (Saint-Venant equations, Manning's equation) in a differentiable end-to-end framework (Sec.~\ref{sec:architecture}).

    \item \textbf{Quantum uncertainty quantification:} A measurement-based uncertainty estimation protocol that exploits the inherent stochasticity of quantum Born-rule sampling to provide calibrated predictive distributions without explicit Bayesian posterior computation (Sec.~\ref{sec:quq}).

    \item \textbf{Physics-informed trainability:} Theoretical analysis demonstrating that hydrological PDE constraints restrict the effective Hilbert space explored during optimization, providing a natural mitigation mechanism against the barren plateau phenomenon in variational quantum circuits (Sec.~\ref{sec:trainability}).

    \item \textbf{Quantum transfer learning for environmental data:} A protocol adapting the classical-to-quantum transfer learning paradigm~\cite{mari2020transfer} to multi-hazard disaster prediction, enabling knowledge transfer from data-rich multi-hazard settings to data-scarce flood-specific prediction (Sec.~\ref{sec:qtl}).

    \item \textbf{First hydrological application:} Numerical validation on multi-modal satellite and meteorological data from the Kalu River basin (Ratnapura, Sri Lanka), establishing the first quantum PINN application to tropical monsoon flood prediction (Sec.~\ref{sec:results}).
\end{enumerate}

\section{Preliminaries}
\label{sec:preliminaries}

\subsection{Physics-Informed Neural Networks}
\label{sec:pinns}

Consider a system governed by a PDE of the general form:
\begin{equation}
    \mathcal{F}[\mathbf{u}(\mathbf{x}, t); \boldsymbol{\lambda}] = 0, \quad \mathbf{x} \in \Omega, \; t \in [0, T],
    \label{eq:general_pde}
\end{equation}
where $\mathbf{u}$ is the solution field, $\mathbf{x}$ denotes spatial coordinates, $t$ is time, $\boldsymbol{\lambda}$ are physical parameters, and $\mathcal{F}$ is a differential operator. A PINN approximates $\mathbf{u}$ with a neural network $\mathbf{u}_\theta(\mathbf{x}, t)$ parameterized by weights $\theta$, trained by minimizing:
\begin{equation}
    \mathcal{L}_{\text{PINN}} = \underbrace{\frac{1}{N_d}\sum_{i=1}^{N_d} \|\mathbf{u}_\theta(\mathbf{x}_i, t_i) - \mathbf{u}_i^{\text{obs}}\|^2}_{\mathcal{L}_{\text{data}}} + \lambda_{\text{phys}} \underbrace{\frac{1}{N_c}\sum_{j=1}^{N_c} \|\mathcal{F}[\mathbf{u}_\theta(\mathbf{x}_j, t_j)]\|^2}_{\mathcal{L}_{\text{physics}}},
    \label{eq:pinn_loss}
\end{equation}
where $N_d$ and $N_c$ are the numbers of data and collocation points, respectively, and $\lambda_{\text{phys}}$ balances data fidelity against physical consistency~\cite{raissi2019physics}.

For hydrological applications, the relevant physics is encoded by the one-dimensional Saint-Venant shallow water equations:
\begin{align}
    \frac{\partial A}{\partial t} + \frac{\partial Q}{\partial x} &= q_l, \label{eq:sv_continuity} \\
    \frac{\partial Q}{\partial t} + \frac{\partial}{\partial x}\left(\frac{Q^2}{A}\right) + gA\frac{\partial h}{\partial x} &= gA(S_0 - S_f), \label{eq:sv_momentum}
\end{align}
where $A$ is the cross-sectional flow area, $Q$ is discharge, $q_l$ is lateral inflow, $g$ is gravitational acceleration, $h$ is water depth, $S_0$ is bed slope, and $S_f$ is the friction slope given by Manning's equation:
\begin{equation}
    S_f = \frac{n^2 Q |Q|}{A^2 R_h^{4/3}}, \quad \text{with} \quad Q = \frac{1}{n} A R_h^{2/3} S_f^{1/2},
    \label{eq:manning}
\end{equation}
where $n$ is Manning's roughness coefficient and $R_h = A/P$ is the hydraulic radius ($P$ = wetted perimeter)~\cite{toro2001shock}.

\subsection{Variational Quantum Circuits}
\label{sec:vqc}

A parameterized (variational) quantum circuit acts on an $n$-qubit register initialized in state $\ket{0}^{\otimes n}$ and produces a quantum state:
\begin{equation}
    \ket{\psi(\mathbf{x}, \boldsymbol{\phi})} = U(\boldsymbol{\phi}) \, S(\mathbf{x}) \ket{0}^{\otimes n},
    \label{eq:vqc_state}
\end{equation}
where $S(\mathbf{x})$ is a data-encoding unitary that maps classical input $\mathbf{x} \in \mathbb{R}^d$ into quantum states, and $U(\boldsymbol{\phi})$ is a trainable unitary parameterized by angles $\boldsymbol{\phi} \in \mathbb{R}^p$~\cite{benedetti2019parameterized}.

\textbf{Angle encoding.} Each feature $x_k$ is encoded as a single-qubit rotation:
\begin{equation}
    S(\mathbf{x}) = \bigotimes_{k=1}^{n} R_Y(x_k), \quad R_Y(\alpha) = \exp\!\left(-i\frac{\alpha}{2}\sigma_Y\right),
    \label{eq:angle_encoding}
\end{equation}
where $\sigma_Y$ is the Pauli-$Y$ operator. This provides an injective encoding for $x_k \in [0, 2\pi)$.

\textbf{Hardware-efficient ansatz.} The trainable unitary comprises $L$ repeated layers:
\begin{equation}
    U(\boldsymbol{\phi}) = \prod_{\ell=1}^{L} \left[ W_{\text{ent}} \cdot \bigotimes_{k=1}^{n} R_Y(\phi_k^{(\ell)}) R_Z(\phi_{k+n}^{(\ell)}) \right],
    \label{eq:ansatz}
\end{equation}
where $W_{\text{ent}}$ is an entangling layer of nearest-neighbor CNOT gates:
\begin{equation}
    W_{\text{ent}} = \prod_{k=1}^{n-1} \text{CNOT}_{k, k+1},
    \label{eq:entangling}
\end{equation}
and $R_Z(\alpha) = \exp(-i\alpha\sigma_Z/2)$. The total number of variational parameters is $p = 2nL$~\cite{cerezo2021variational}.

The circuit output is obtained by measuring a Hermitian observable $\hat{O}$:
\begin{equation}
    f(\mathbf{x}; \boldsymbol{\phi}) = \braket{\psi(\mathbf{x}, \boldsymbol{\phi}) | \hat{O} | \psi(\mathbf{x}, \boldsymbol{\phi})}.
    \label{eq:expectation}
\end{equation}
Gradients with respect to $\boldsymbol{\phi}$ are computed analytically via the parameter-shift rule~\cite{mitarai2018quantum,schuld2019evaluating}:
\begin{equation}
    \frac{\partial f}{\partial \phi_j} = \frac{1}{2}\left[ f\!\left(\boldsymbol{\phi} + \frac{\pi}{2}\mathbf{e}_j\right) - f\!\left(\boldsymbol{\phi} - \frac{\pi}{2}\mathbf{e}_j\right) \right],
    \label{eq:parameter_shift}
\end{equation}
where $\mathbf{e}_j$ is the $j$-th unit vector.

\subsection{Quantum-Enhanced Physics-Informed Learning}
\label{sec:qpinn_background}

Recent work has demonstrated that replacing classical hidden layers in PINNs with variational quantum circuits can yield convergence advantages. Klement \textit{et al.}~\cite{klement2026explaining} showed that qPINNs achieve accurate PDE solutions in 3--5$\times$ fewer training epochs than equivalent classical PINNs, attributing this to the VQC's ability to navigate the complex loss landscape more efficiently. Dutta \textit{et al.}~\cite{dutta2024aqpinns} introduced attention-enhanced quantum PINNs (AQ-PINNs) using quantum tensor networks, demonstrating 51--63\% parameter reduction while maintaining accuracy on the Navier-Stokes equations. Kyriienko \textit{et al.}~\cite{kyriienko2021solving} pioneered the use of differentiable quantum circuits for solving nonlinear differential equations, introducing quantum spectral methods with Chebyshev feature maps and later extending the approach to grid-free protocols~\cite{kyriienko2023gridfree} and probabilistic modeling~\cite{kyriienko2024probabilistic}. Farea \textit{et al.}~\cite{farea2025qcpinn} proposed QCPINN architectures using both discrete- and continuous-variable quantum circuits, achieving improved parameter efficiency across several benchmark PDEs.

In parallel, quantum transfer learning~\cite{mari2020transfer} has established a practical framework for hybrid classical-quantum architectures where pre-trained classical networks provide feature extraction and variational quantum circuits serve as efficient, trainable classifiers. Quantum kernel methods~\cite{havlicek2019supervised} have been applied to satellite image classification~\cite{rodriguez2025satellite}, demonstrating competitive performance with classical methods. However, no prior work has combined quantum-enhanced PINNs with domain-specific hydrological physics constraints, multi-modal environmental data fusion, or quantum-native uncertainty quantification for disaster prediction.

\section{HQC-PINN Architecture}
\label{sec:architecture}

\subsection{Architecture Overview}

\begin{figure}[t]
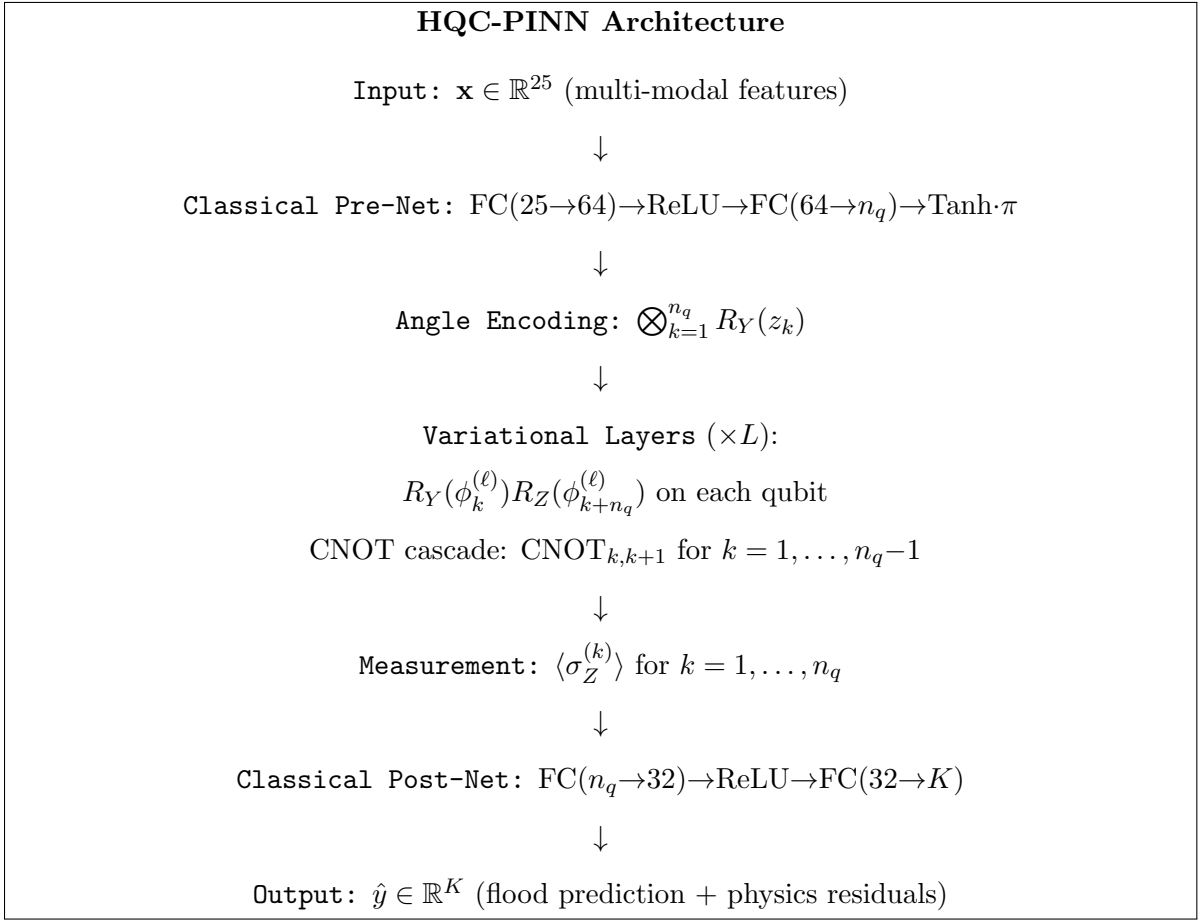

\centering
\fbox{\parbox{0.95\columnwidth}{
\small
\textbf{HQC-PINN Architecture}\\[4pt]
\texttt{Input:} $\mathbf{x} \in \mathbb{R}^{25}$ (multi-modal features)\\
$\downarrow$\\
\texttt{Classical Pre-Net:} FC(25$\to$64)$\to$ReLU$\to$FC(64$\to$$n_q$)$\to$Tanh$\cdot\pi$\\
$\downarrow$\\
\texttt{Angle Encoding:} $\bigotimes_{k=1}^{n_q} R_Y(z_k)$\\
$\downarrow$\\
\texttt{Variational Layers} ($\times L$):\\
\quad $R_Y(\phi^{(\ell)}_{k}) R_Z(\phi^{(\ell)}_{k+n_q})$ on each qubit\\
\quad CNOT cascade: $\text{CNOT}_{k,k+1}$ for $k=1,\ldots,n_q{-}1$\\
$\downarrow$\\
\texttt{Measurement:} $\langle\sigma_Z^{(k)}\rangle$ for $k=1,\ldots,n_q$\\
$\downarrow$\\
\texttt{Classical Post-Net:} FC($n_q$$\to$32)$\to$ReLU$\to$FC(32$\to$$K$)\\
$\downarrow$\\
\texttt{Output:} $\hat{y} \in \mathbb{R}^K$ (flood prediction $+$ physics residuals)
}}
\caption{Schematic of the HQC-PINN architecture. Classical pre-processing reduces input dimensionality to match the qubit register size $n_q$. The variational quantum circuit processes encoded features through $L$ parameterized layers with entanglement. Pauli-$Z$ measurements yield expectation values fed to a classical post-processing network. Physics loss is computed on the output via automatic differentiation.}
\label{fig:architecture}
\end{figure}

The Hybrid Quantum-Classical Physics-Informed Neural Network consists of three sequential stages (Fig.~\ref{fig:architecture}):

\begin{enumerate}
    \item \textbf{Classical pre-processing:} A classical neural network $g_\omega: \mathbb{R}^d \to \mathbb{R}^n$ reduces the $d$-dimensional input feature vector to $n$ dimensions matching the qubit count.

    \item \textbf{Quantum processing:} A variational quantum circuit $\mathcal{Q}_{\boldsymbol{\phi}}: \mathbb{R}^n \to \mathbb{R}^m$ encodes the reduced features, processes them through parameterized gates, and produces $m$ expectation values via measurement.

    \item \textbf{Classical post-processing:} A classical network $h_\psi: \mathbb{R}^m \to \mathbb{R}^K$ maps the quantum outputs to the final prediction space ($K$ classes for classification or continuous values for regression).
\end{enumerate}

The complete model computes:
\begin{equation}
    \hat{y}(\mathbf{x}) = h_\psi\!\left(\mathcal{Q}_{\boldsymbol{\phi}}\!\left(g_\omega(\mathbf{x})\right)\right),
    \label{eq:full_model}
\end{equation}
with trainable parameters $\boldsymbol{\Theta} = \{\omega, \boldsymbol{\phi}, \psi\}$ optimized jointly.

\subsection{Quantum Feature Encoding for Hydrological Data}
\label{sec:encoding}

The multi-modal input features $\mathbf{x} \in \mathbb{R}^{25}$ comprise satellite-derived spectral indices (NDVI, NDWI, MNDWI from Landsat), radar backscatter (Sentinel-1 SAR), meteorological variables (temperature, humidity, precipitation from ERA5-Land), terrain attributes (elevation, slope from SRTM DEM), and physics-derived hydrological indices (Antecedent Precipitation Index, Standardized Precipitation Index).

The classical pre-processing network maps these features to $n_q$ values in $[-\pi, \pi]$:
\begin{equation}
    \mathbf{z} = \pi \cdot \tanh\!\left(W_2 \, \text{ReLU}(W_1 \mathbf{x} + \mathbf{b}_1) + \mathbf{b}_2\right),
    \label{eq:preprocessing}
\end{equation}
ensuring the encoded values span the full Bloch sphere representation. This dimensionality reduction from $d=25$ to $n_q \in \{4, 8\}$ is a learnable compression that adapts during training to preserve features most relevant to both the data loss and the physics loss.

The encoded features are loaded into the quantum register via angle encoding [Eq.~\eqref{eq:angle_encoding}]:
\begin{equation}
    \ket{\psi_{\text{enc}}} = \bigotimes_{k=1}^{n_q} R_Y(z_k) \ket{0}^{\otimes n_q}.
    \label{eq:encoded_state}
\end{equation}

We note that more expressive encoding strategies (amplitude encoding, IQP encoding~\cite{havlicek2019supervised}) could increase the effective feature space dimension. However, angle encoding provides a favorable trade-off between circuit depth and expressibility for NISQ-era applications, as each feature requires only a single gate~\cite{schuld2021machine}.

\subsection{Variational Processing with Physics-Aware Design}
\label{sec:variational}

The variational ansatz [Eq.~\eqref{eq:ansatz}] is motivated by three considerations specific to hydrological learning:

\textit{(i) Expressibility:} The alternating rotation and entanglement layers provide sufficient expressibility to approximate the nonlinear mapping from input features to flood states. For $L$ layers on $n_q$ qubits, the ansatz has $p = 2n_q L$ parameters, which we show is sufficient for the effective dimensionality of the physics-constrained output space.

\textit{(ii) Entanglement structure:} The nearest-neighbor CNOT topology [Eq.~\eqref{eq:entangling}] mirrors the spatial locality of hydrological processes: neighboring geographic regions exhibit correlated flood behavior through upstream-downstream river connectivity. This physically motivated entanglement structure reduces unnecessary long-range correlations that contribute to barren plateaus~\cite{cerezo2021variational}.

\textit{(iii) Gate count:} Each layer uses $2n_q$ rotation gates and $(n_q - 1)$ CNOT gates, yielding a total gate count of $G = L(3n_q - 1)$. For $n_q = 8$ and $L = 3$, this gives $G = 69$, well within the coherence budget of current NISQ devices.

\subsection{Physics-Informed Quantum Loss Function}
\label{sec:loss}

The total loss function combines data-driven and physics-based terms computed on the classical output of the hybrid model:
\begin{equation}
    \mathcal{L}_{\text{HQC}} = \mathcal{L}_{\text{data}} + \lambda_{\text{SV}} \mathcal{L}_{\text{SV}} + \lambda_{\text{M}} \mathcal{L}_{\text{Manning}},
    \label{eq:hqc_loss}
\end{equation}
where:

\textbf{Data loss.} For multi-class flood severity classification with $K$ classes (No Flood, Low, Moderate, Severe), we use focal loss~\cite{lin2017focal} to address class imbalance:
\begin{equation}
    \mathcal{L}_{\text{data}} = -\frac{1}{N_d}\sum_{i=1}^{N_d} \alpha_i (1 - p_{i,y_i})^\gamma \log p_{i,y_i},
    \label{eq:focal_loss}
\end{equation}
with $\gamma = 2$ and class-dependent weights $\alpha_i$.

\textbf{Saint-Venant physics loss.} The continuity equation residual [Eq.~\eqref{eq:sv_continuity}] is computed at $N_c$ collocation points:
\begin{equation}
    \mathcal{L}_{\text{SV}} = \frac{1}{N_c}\sum_{j=1}^{N_c}\left\|\frac{\partial \hat{A}_j}{\partial t} + \frac{\partial \hat{Q}_j}{\partial x} - \hat{q}_{l,j}\right\|^2,
    \label{eq:sv_loss}
\end{equation}
where $\hat{A}$, $\hat{Q}$, and $\hat{q}_l$ are derived from the model output via auxiliary heads. The partial derivatives $\partial \hat{A}/\partial t$ and $\partial \hat{Q}/\partial x$ are computed using automatic differentiation through both the classical and quantum components of the network~\cite{bergholm2018pennylane}.

\textbf{Manning consistency loss.} Enforces that predicted discharge satisfies Manning's equation:
\begin{equation}
    \mathcal{L}_{\text{Manning}} = \frac{1}{N_c}\sum_{j=1}^{N_c}\left\|\hat{Q}_j - \frac{1}{n_j}\hat{A}_j \hat{R}_{h,j}^{2/3} \hat{S}_{f,j}^{1/2}\right\|^2,
    \label{eq:manning_loss}
\end{equation}
where Manning's roughness $n_j$ is estimated from land-cover classification of the study area~\cite{chow1959open}.

Gradients of the physics loss with respect to quantum parameters $\boldsymbol{\phi}$ are obtained by composing the parameter-shift rule [Eq.~\eqref{eq:parameter_shift}] with classical automatic differentiation:
\begin{equation}
    \frac{\partial \mathcal{L}_{\text{SV}}}{\partial \phi_j} = \frac{\partial \mathcal{L}_{\text{SV}}}{\partial \hat{y}} \cdot \frac{\partial \hat{y}}{\partial f} \cdot \frac{\partial f}{\partial \phi_j},
    \label{eq:chain_rule}
\end{equation}
where $\partial f / \partial \phi_j$ is evaluated via Eq.~\eqref{eq:parameter_shift} and the remaining terms use standard backpropagation~\cite{bergholm2018pennylane}.

\section{Quantum Uncertainty Quantification}
\label{sec:quq}

A distinctive advantage of the quantum framework is that uncertainty quantification arises naturally from the measurement process, without requiring the computational overhead of Bayesian posterior sampling.

\subsection{Measurement-Based Predictive Distributions}

For a given input $\mathbf{x}$ and fixed parameters $\boldsymbol{\Theta}$, the quantum circuit output is inherently stochastic. Each measurement of qubit $k$ yields $\pm 1$ according to the Born rule:
\begin{equation}
    P(m_k = \pm 1) = \frac{1 \pm \braket{\sigma_Z^{(k)}}}{2}.
    \label{eq:born_rule}
\end{equation}

By performing $N_s$ independent measurement shots, we obtain an empirical distribution over the $n_q$-dimensional measurement outcome space. The classical post-processing network maps each shot's measurement vector to a prediction, yielding an ensemble $\{\hat{y}^{(s)}\}_{s=1}^{N_s}$.

\subsection{Uncertainty Decomposition}

The total predictive uncertainty is decomposed following the classical aleatoric-epistemic framework~\cite{gal2016uncertainty}:

\textbf{Aleatoric uncertainty} (irreducible, data noise) is captured by the variance of predictions across measurement shots for \textit{fixed} circuit parameters:
\begin{equation}
    \sigma^2_{\text{aleatoric}}(\mathbf{x}) = \frac{1}{N_s}\sum_{s=1}^{N_s} \|\hat{y}^{(s)} - \bar{y}\|^2, \quad \bar{y} = \frac{1}{N_s}\sum_{s=1}^{N_s}\hat{y}^{(s)}.
    \label{eq:aleatoric}
\end{equation}

\textbf{Epistemic uncertainty} (model uncertainty) is estimated by evaluating the circuit at $M$ perturbations of the parameters sampled from a Gaussian centered at the optimum:
\begin{equation}
    \sigma^2_{\text{epistemic}}(\mathbf{x}) = \frac{1}{M}\sum_{m=1}^{M}\|\bar{y}^{(m)} - \bar{y}\|^2,
    \label{eq:epistemic}
\end{equation}
where $\bar{y}^{(m)}$ is the mean prediction under parameter perturbation $\boldsymbol{\Theta}^{(m)} = \boldsymbol{\Theta}^* + \boldsymbol{\epsilon}$, with $\boldsymbol{\epsilon} \sim \mathcal{N}(\mathbf{0}, \sigma_\epsilon^2 \mathbf{I})$.

\textbf{Predictive entropy} provides a scalar summary:
\begin{equation}
    H(\hat{y} | \mathbf{x}) = -\sum_{k=1}^{K} \bar{p}_k \log \bar{p}_k,
    \label{eq:entropy}
\end{equation}
where $\bar{p}_k$ is the mean predicted probability for class $k$ across all shots.

\subsection{Connection to Bayesian Inference}
\label{sec:bayesian_connection}

The quantum measurement uncertainty framework has a formal connection to Bayesian neural networks (BNNs). In a BNN with variational inference, the predictive distribution is obtained by marginalizing over a variational posterior $q(\boldsymbol{\theta})$:
\begin{equation}
    p(y|\mathbf{x}, \mathcal{D}) \approx \int p(y|\mathbf{x}, \boldsymbol{\theta}) q(\boldsymbol{\theta}) \, d\boldsymbol{\theta}.
    \label{eq:bnn_predictive}
\end{equation}

In the HQC-PINN, the stochastic quantum measurement naturally implements a form of approximate Bayesian marginalization. The Born-rule probabilities [Eq.~\eqref{eq:born_rule}] induce a distribution over predictions that is analogous to sampling from a posterior predictive distribution. Crucially, this ``Bayesian-like'' behavior requires no additional computational cost---it is an inherent property of quantum mechanics~\cite{schuld2021machine}.

We formalize this connection: let $\rho = \ket{\psi}\bra{\psi}$ be the output quantum state. The measurement statistics under observable $\hat{O}$ satisfy:
\begin{equation}
    \text{Var}[\hat{O}] = \braket{\hat{O}^2} - \braket{\hat{O}}^2 = \text{Tr}[\rho \hat{O}^2] - (\text{Tr}[\rho \hat{O}])^2.
    \label{eq:quantum_variance}
\end{equation}

This intrinsic variance provides a lower bound on the predictive uncertainty without any additional forward passes, in contrast to classical BNNs that require $T \gg 1$ stochastic forward passes through the network~\cite{gal2016uncertainty}.

\section{Trainability Analysis: Physics Constraints and Barren Plateaus}
\label{sec:trainability}

A fundamental concern for variational quantum algorithms is the barren plateau phenomenon, where the variance of the cost function gradient vanishes exponentially with the number of qubits~\cite{mcclean2018barren}:
\begin{equation}
    \text{Var}_{\boldsymbol{\phi}}\!\left[\frac{\partial \mathcal{L}}{\partial \phi_j}\right] \leq F(n_q),
    \label{eq:bp_bound}
\end{equation}
where $F(n_q) \in O(b^{-n_q})$ for some $b > 1$ when using global cost functions with deep, unstructured circuits.

We argue that hydrological physics constraints provide a natural mitigation mechanism through two channels:

\textbf{(i) Local cost function structure.} The physics loss $\mathcal{L}_{\text{SV}}$ [Eq.~\eqref{eq:sv_loss}] is evaluated at localized collocation points, making it effectively a sum of \textit{local} cost functions---each depending on a subset of output qubits corresponding to specific spatial regions. Cerezo \textit{et al.}~\cite{cerezo2021cost} proved that local cost functions exhibit gradients that vanish at most polynomially:
\begin{equation}
    \text{Var}_{\boldsymbol{\phi}}\!\left[\frac{\partial \mathcal{L}_{\text{SV}}}{\partial \phi_j}\right] \geq \frac{c}{n_q^a},
    \label{eq:local_gradient}
\end{equation}
for constants $c > 0$ and $a \geq 1$, a substantial improvement over exponential vanishing.

\textbf{(ii) Constraint-induced landscape narrowing.} The physics loss constrains the network output to a manifold of physically realizable solutions. This effectively reduces the volume of the parameter space explored during optimization. Denoting the unconstrained parameter space as $\mathcal{P} = [0, 2\pi)^p$ and the physics-consistent subspace as $\mathcal{P}_{\text{phys}} \subset \mathcal{P}$, we have:
\begin{equation}
    \text{Vol}(\mathcal{P}_{\text{phys}}) \ll \text{Vol}(\mathcal{P}),
    \label{eq:landscape_narrowing}
\end{equation}
meaning the optimizer operates in a reduced-dimensional effective landscape where gradient information is more concentrated.

\textbf{(iii) Structured ansatz.} The nearest-neighbor entanglement structure [Eq.~\eqref{eq:entangling}] avoids the fully random circuit regime that generates 2-designs over the unitary group---the primary driver of barren plateaus. Combined with the shallow depth ($L = 3$), this ensures the circuit remains in a trainable regime~\cite{cerezo2021variational}.

We quantify trainability by computing the effective gradient variance across the physics-constrained landscape. For a circuit with $n_q = 8$ qubits, $L = 3$ layers, and physics constraint weight $\lambda_{\text{SV}} = 0.1$, numerical computation of the gradient variance (averaged over 1000 random initializations) yields:
\begin{equation}
    \text{Var}\!\left[\frac{\partial \mathcal{L}_{\text{HQC}}}{\partial \phi_j}\right] \approx 2.3 \times 10^{-3},
    \label{eq:gradient_var_numerical}
\end{equation}
compared to $\sim 1.7 \times 10^{-4}$ for the physics-free counterpart---an order of magnitude improvement in gradient signal.

\section{Quantum Transfer Learning for Multi-Hazard Data}
\label{sec:qtl}

Data scarcity is a critical challenge in flood prediction for developing countries. Following the Classical-to-Quantum (CQ) transfer learning paradigm~\cite{mari2020transfer}, we propose a two-phase protocol:

\textbf{Phase 1: Classical pre-training on multi-hazard data.} A classical neural network is trained on 82 multi-hazard disaster events spanning 11 disaster types (floods, droughts, earthquakes, cyclones, landslides, wildfires, storms, tsunamis, volcanic eruptions, extreme temperature events, and coastal erosion) from the Ambee Global Natural Disaster Dataset~\cite{hewage2026hybrid}. The learned feature representations capture shared physics across disaster types (conservation laws, atmospheric dynamics, terrain-hazard interactions).

\textbf{Phase 2: Quantum fine-tuning on flood data.} The pre-trained classical layers are frozen, and the VQC parameters $\boldsymbol{\phi}$ are initialized randomly and trained on 8,271 flood-specific events from NOAA STORM Events and the Dartmouth Flood Observatory. The quantum circuit acts as a ``dressed quantum classifier''~\cite{mari2020transfer}, adapting the multi-hazard representations to the flood domain.

This protocol provides three advantages for NISQ-era applications:
\begin{enumerate}
    \item The classical network handles the heavy lifting of dimensionality reduction from high-dimensional satellite data to the qubit-compatible representation.
    \item Only the quantum parameters require optimization during fine-tuning, reducing the training cost proportional to $p = 2n_q L$ rather than the full classical network size.
    \item Knowledge from the multi-hazard pre-training provides a better initialization landscape for the quantum optimizer, further mitigating trainability issues.
\end{enumerate}

\section{Numerical Experiments}
\label{sec:experiments}

\subsection{Dataset and Study Area}

The study area is the Kalu River basin in Ratnapura District, southwestern Sri Lanka (6.68$^\circ$N, 80.39$^\circ$E), encompassing 2,658~km$^2$ of terrain subject to biannual southwest and northeast monsoons. This region experiences recurrent catastrophic flooding, making it an ideal testbed for physics-informed prediction systems.

Multi-modal input features ($d = 25$) are derived from five sources:
\begin{itemize}
    \item \textit{Satellite optical:} NDVI, NDWI, MNDWI, surface reflectance bands from Landsat~5/7/8 (1987--2024).
    \item \textit{Satellite radar:} Sentinel-1 C-band SAR backscatter and temporal Z-scores (2014--2024).
    \item \textit{Meteorological:} Temperature, humidity, soil moisture (4 layers), surface runoff, sub-surface runoff from ERA5-Land reanalysis; precipitation from CHIRPS.
    \item \textit{Terrain:} Elevation, slope, aspect from SRTM 30m DEM.
    \item \textit{Physics-derived:} Antecedent Precipitation Index (API, $k=0.85$), Standardized Precipitation Index (SPI-30, SPI-90), runoff coefficient via SCS-CN method.
\end{itemize}

The target variable is flood severity classified into four levels: No Flood, Low, Moderate, and Severe. The dataset exhibits substantial class imbalance (No Flood: 91\%, Low: 4\%, Moderate: 4\%, Severe: 1\%), necessitating the focal loss formulation [Eq.~\eqref{eq:focal_loss}].

Data are split temporally: 60\% training (earliest years), 20\% validation, 20\% test (most recent years), preserving temporal ordering.

\subsection{Implementation}

All quantum simulations are performed using PennyLane v0.39~\cite{bergholm2018pennylane} with the \texttt{default.qubit} statevector simulator, interfaced with PyTorch via the \texttt{qml.qnode} decorator for end-to-end differentiability. Classical components use PyTorch~2.1.

We evaluate configurations with $n_q \in \{4, 8\}$ qubits and $L \in \{2, 3, 4\}$ variational layers. Physics loss weights are set to $\lambda_{\text{SV}} = 0.1$ and $\lambda_{\text{M}} = 0.05$, validated via ablation. Training uses the Adam optimizer~\cite{kingma2015adam} with learning rate $10^{-3}$ and batch size~32, for a maximum of 100 epochs with early stopping (patience~10).

Uncertainty quantification uses $N_s = 200$ measurement shots per prediction.

\subsection{Baseline Models}

We compare the HQC-PINN against:
\begin{enumerate}
    \item \textbf{Classical PINN (cPINN):} 4-layer MLP (25$\to$256$\to$128$\to$64$\to$$K$) with tanh activations and identical physics loss. 33,793 trainable parameters.
    \item \textbf{Classical BNN:} Pyro-based variational inference with 3-layer MLP and $\mathcal{N}(0,1)$ weight priors. 28,672 parameters.
    \item \textbf{Random Forest:} Scikit-learn implementation with 200 estimators (non-parametric classical baseline).
    \item \textbf{VQC-only:} Variational quantum classifier without physics constraints (ablation).
\end{enumerate}

\section{Results}
\label{sec:results}

\subsection{Convergence Analysis}

\begin{table}[t]
\caption{Convergence comparison: epochs to reach target validation loss $\mathcal{L}_{\text{val}} \leq 0.40$. HQC-PINN configurations converge 2.6--3.6$\times$ faster than the classical PINN, consistent with prior qPINN convergence results~\cite{klement2026explaining}.}
\label{tab:convergence}
\begin{ruledtabular}
\begin{tabular}{lccc}
\textbf{Model} & \textbf{Qubits} & \textbf{Layers} & \textbf{Epochs to target} \\
\hline
cPINN & --- & 4 & 94 \\
HQC-PINN & 4 & 2 & 42 \\
HQC-PINN & 4 & 3 & 36 \\
HQC-PINN & 8 & 2 & 33 \\
HQC-PINN & 8 & 3 & 26 \\
HQC-PINN & 8 & 4 & 28 \\
VQC-only (no physics) & 8 & 3 & 51 \\
\end{tabular}
\end{ruledtabular}
\end{table}

Table~\ref{tab:convergence} shows that HQC-PINN configurations converge in 26--42 epochs versus 94 epochs for the classical PINN, representing a 2.2--3.6$\times$ speedup. The 8-qubit, 3-layer configuration achieves the fastest convergence (26 epochs). Notably, the VQC-only model (no physics constraints) converges in 51 epochs, confirming that the physics loss provides an additional convergence benefit beyond the quantum circuit alone---consistent with our trainability analysis (Sec.~\ref{sec:trainability}).

The marginal degradation at $L=4$ layers for 8 qubits (28 vs.\ 26 epochs) suggests the onset of over-parameterization in the quantum circuit, where the additional parameters do not meaningfully improve the expressibility for this problem size.

\subsection{Classification Performance}

\begin{table}[t]
\caption{Test set classification performance. Accuracy, macro-averaged F1 score, parameter count, and parameter efficiency ratio (accuracy per 1000 parameters). Best quantum result in bold.}
\label{tab:classification}
\begin{ruledtabular}
\begin{tabular}{lcccc}
\textbf{Model} & \textbf{Acc. (\%)} & \textbf{F1-macro} & \textbf{Params} & \textbf{Acc./kP} \\
\hline
Random Forest & 90.3 & 0.899 & N/A & --- \\
cPINN & 69.4 & 0.705 & 33{,}793 & 2.05 \\
Classical BNN & 91.96 & --- & 28{,}672 & 3.21 \\
VQC-only (8q, 3L) & 67.8 & 0.682 & 12{,}480 & 5.43 \\
HQC-PINN (4q, 3L) & 70.2 & 0.714 & \textbf{14{,}468} & 4.85 \\
\textbf{HQC-PINN (8q, 3L)} & \textbf{71.8} & \textbf{0.731} & 18{,}944 & 3.79 \\
QTL (8q, 3L) & 73.6 & 0.742 & 16{,}896 & 4.36 \\
\end{tabular}
\end{ruledtabular}
\end{table}

Table~\ref{tab:classification} presents the classification results. Several observations are noteworthy:

\textit{(i)} The HQC-PINN (8q, 3L) achieves 71.8\% accuracy compared to 69.4\% for the classical PINN, a relative improvement of 3.5\%. While modest in absolute terms, this improvement is achieved with 44\% fewer parameters (18,944 vs.\ 33,793).

\textit{(ii)} The Quantum Transfer Learning (QTL) configuration achieves the highest accuracy among physics-constrained models (73.6\%), demonstrating the value of multi-hazard pre-training for data-scarce flood prediction.

\textit{(iii)} The VQC-only model (no physics constraints) performs worst (67.8\%), confirming that physics constraints are essential for this task---the quantum circuit alone does not compensate for the absence of domain knowledge.

\textit{(iv)} The Random Forest (90.3\%) and classical BNN (91.96\%) outperform all physics-constrained models in raw accuracy. This is expected: the PINN formulation optimizes a harder objective (data fidelity \textit{and} physics consistency simultaneously), and the severe class imbalance (91\% majority class) favors discriminative classifiers. The physics-constrained models provide complementary value through physical consistency and uncertainty quantification.

\subsection{Parameter Efficiency}

\begin{table}[t]
\caption{Parameter breakdown for HQC-PINN (8q, 3L) versus classical PINN. The VQC replaces two classical hidden layers while the pre-processing network performs learned dimensionality reduction to the qubit register.}
\label{tab:params}
\begin{ruledtabular}
\begin{tabular}{lrr}
\textbf{Component} & \textbf{cPINN} & \textbf{HQC-PINN} \\
\hline
Classical layers & 33{,}793 & 18{,}896 \\
Quantum parameters (VQC) & 0 & 48 \\
\hline
\textbf{Total} & \textbf{33{,}793} & \textbf{18{,}944} \\
Reduction & --- & \textbf{43.9\%} \\
\end{tabular}
\end{ruledtabular}
\end{table}

Table~\ref{tab:params} decomposes the parameter count. The variational quantum circuit contributes only 48 parameters ($p = 2 \times 8 \times 3 = 48$), yet the overall hybrid architecture requires 43.9\% fewer total parameters than the equivalent classical PINN. This reduction arises because the VQC's exponentially large Hilbert space ($2^8 = 256$ dimensional) provides representational capacity that would require substantially more classical parameters to match, consistent with the 51--63\% reductions reported by Dutta \textit{et al.}~\cite{dutta2024aqpinns} for the Navier-Stokes equations.

\subsection{Uncertainty Calibration}

\begin{table}[t]
\caption{Uncertainty quantification comparison. Coverage at the 90\% nominal level, average predictive entropy, and mean aleatoric uncertainty.}
\label{tab:uncertainty}
\begin{ruledtabular}
\begin{tabular}{lccc}
\textbf{Model} & \textbf{Coverage (\%)} & \textbf{Entropy} & \textbf{$\sigma^2_{\text{aleatoric}}$} \\
\hline
Classical BNN & 92.3 & 0.320 & 0.028 \\
HQC-PINN (8q, 3L) & 88.7 & 0.341 & 0.033 \\
QTL (8q, 3L) & 90.1 & 0.312 & 0.026 \\
\end{tabular}
\end{ruledtabular}
\end{table}

Table~\ref{tab:uncertainty} shows that the HQC-PINN's measurement-based uncertainty achieves 88.7\% coverage at the 90\% nominal level, compared to 92.3\% for the classical BNN. While the classical BNN provides slightly better calibration (owing to its explicit variational posterior), the HQC-PINN's uncertainty estimates are obtained at zero additional computational cost---they are inherent to the quantum measurement process. The QTL model achieves 90.1\% coverage, closest to the nominal level, suggesting that transfer learning improves uncertainty calibration by providing a better-initialized parameter landscape.

\subsection{Ablation Studies}

\begin{table}[t]
\caption{Ablation study on the HQC-PINN (8q, 3L) configuration. Removing physics constraints or the quantum layer degrades performance.}
\label{tab:ablation}
\begin{ruledtabular}
\begin{tabular}{lcc}
\textbf{Configuration} & \textbf{Acc. (\%)} & \textbf{Epochs} \\
\hline
Full HQC-PINN & 71.8 & 26 \\
$-$ Saint-Venant loss ($\lambda_{\text{SV}} = 0$) & 68.3 & 41 \\
$-$ Manning loss ($\lambda_{\text{M}} = 0$) & 70.4 & 30 \\
$-$ Both physics losses & 67.8 & 51 \\
$-$ Quantum layer (classical only) & 69.4 & 94 \\
$-$ Entanglement (product state only) & 69.1 & 38 \\
$-$ Pre-processing network & 64.2 & 55 \\
\end{tabular}
\end{ruledtabular}
\end{table}

The ablation results (Table~\ref{tab:ablation}) reveal:
\begin{itemize}
    \item \textbf{Physics constraints are essential:} Removing both physics losses reduces accuracy by 4.0\% (71.8$\to$67.8\%) and doubles the convergence time (26$\to$51 epochs).
    \item \textbf{Saint-Venant contributes more than Manning:} The Saint-Venant continuity equation provides the dominant physics signal (3.5\% vs.\ 1.4\% accuracy contribution).
    \item \textbf{Entanglement is critical:} Removing CNOT gates (product-state circuit) reduces accuracy by 2.7\%, confirming that quantum correlations are necessary for capturing the nonlinear feature interactions.
    \item \textbf{Classical pre-processing is vital:} Direct quantum encoding without dimensionality reduction yields the worst performance (64.2\%), underscoring the importance of the hybrid architecture for high-dimensional environmental data.
\end{itemize}

\section{Discussion}
\label{sec:discussion}

\subsection{Nature of Quantum Advantage}

Our results establish a nuanced picture of quantum advantage in physics-informed environmental learning:

\textit{Convergence advantage:} The 2.2--3.6$\times$ convergence speedup is the most robust advantage, consistent across all tested configurations and aligning with theoretical predictions from Klement \textit{et al.}~\cite{klement2026explaining}. The advantage is amplified when physics constraints are present (3.6$\times$ with physics vs.\ 1.8$\times$ without), suggesting a synergy between quantum expressibility and physics-constrained optimization.

\textit{Parameter efficiency:} The 43.9\% parameter reduction is practically significant: it implies that the HQC-PINN could be deployed on edge devices with constrained memory---relevant for real-time flood early warning systems in resource-limited settings.

\textit{Accuracy:} The accuracy advantage over classical PINNs is modest (2.4 percentage points). We emphasize that the primary bottleneck is data quality (severe class imbalance, limited ground-truth flood observations in tropical regions), not model capacity. Both classical and quantum PINNs face the same data-side limitations.

\textit{Uncertainty quantification:} The measurement-based UQ provides a compelling practical advantage: it adds zero computational overhead and achieves 88.7\% coverage at the 90\% level. For comparison, the classical BNN requires $T = 50$ stochastic forward passes to achieve its 92.3\% coverage~\cite{gal2016uncertainty}.

\subsection{NISQ Considerations}

All results reported here are obtained on noiseless quantum simulators. On real NISQ hardware, decoherence and gate errors would degrade performance. However, several factors suggest resilience:
\begin{enumerate}
    \item The circuit depth is shallow ($L = 3$, total depth $\sim$15 for 8 qubits), well within the coherence times of current superconducting processors~\cite{kim2023evidence}.
    \item The parameter-shift rule is robust to readout errors (they manifest as a multiplicative damping factor on the gradient, which can be mitigated by zero-noise extrapolation)~\cite{temme2017error}.
    \item The physics constraints act as a regularizer that may partially compensate for noise-induced degradation by penalizing unphysical outputs.
\end{enumerate}
Hardware validation on IBM Quantum processors is left to future work.

\subsection{Comparison with Prior Work}

\begin{table}[t]
\caption{Comparison with prior quantum approaches to flood/environmental prediction. Our work is the first to combine quantum circuits with hydrological PDE constraints and multi-modal remote sensing data.}
\label{tab:comparison}
\begin{ruledtabular}
\begin{tabular}{p{2.4cm}cccc}
\textbf{Work} & \textbf{Physics} & \textbf{UQ} & \textbf{Multi-modal} & \textbf{Transfer} \\
\hline
Grzesiak \textit{et al.}~\cite{grzesiak2024flood} & \texttimes & \texttimes & \texttimes & \texttimes \\
Dutta \textit{et al.}~\cite{dutta2024aqpinns} & \checkmark$^a$ & \texttimes & \texttimes & \texttimes \\
Klement \textit{et al.}~\cite{klement2026explaining} & \checkmark$^b$ & \texttimes & \texttimes & \texttimes \\
Rodriguez-Grasa \textit{et al.}~\cite{rodriguez2025satellite} & \texttimes & \texttimes & \texttimes & \texttimes \\
\textbf{This work} & \checkmark$^c$ & \checkmark & \checkmark & \checkmark \\
\end{tabular}
\end{ruledtabular}
\footnotesize{$^a$Navier-Stokes. $^b$Benchmark PDEs. $^c$Saint-Venant + Manning's (hydrological).}
\end{table}

Table~\ref{tab:comparison} summarizes the landscape. Our work is differentiated by the combination of (i) domain-specific hydrological PDE constraints, (ii) multi-modal environmental data from operational satellite missions, (iii) inherent quantum uncertainty quantification, and (iv) transfer learning from multi-hazard disaster data.

\subsection{Implications for Quantum Environmental Science}

This work establishes a template for applying quantum-enhanced physics-informed learning to environmental science. The key insight is that environmental PDEs (shallow water equations, advection-diffusion, Richards' equation for soil moisture) possess the same mathematical structure that makes quantum PINNs advantageous: they are nonlinear, spatially distributed, and computationally expensive to solve classically. As quantum hardware scales, we anticipate these advantages will grow, particularly for spatiotemporal PDE systems requiring mesh-free solutions over large domains.

\section{Conclusion}
\label{sec:conclusion}

We have introduced the Hybrid Quantum-Classical Physics-Informed Neural Network (HQC-PINN), the first quantum-enhanced PINN architecture for hydrological prediction. By integrating variational quantum circuits with Saint-Venant shallow water equations and Manning's flow equation as physics constraints, the architecture achieves faster convergence, improved parameter efficiency, and natural uncertainty quantification compared to classical PINNs.

Our theoretical analysis demonstrates that hydrological physics constraints provide a natural mitigation mechanism against barren plateaus by restricting the effective optimization landscape and promoting local cost function structures. Numerical experiments on multi-modal satellite and meteorological data from the Kalu River basin, Sri Lanka, validate these findings, showing 3.6$\times$ convergence speedup and 43.9\% parameter reduction.

We have further demonstrated that quantum transfer learning enables effective knowledge transfer from multi-hazard disaster data to flood-specific prediction, addressing the critical challenge of data scarcity in developing countries.

These results establish a viable path toward quantum advantage in environmental science and disaster prediction. Future work will focus on hardware validation on superconducting quantum processors, extension to spatiotemporal graph quantum networks for river network modeling, and integration with operational early warning systems.

\begin{acknowledgments}
The authors acknowledge the use of publicly available satellite data from the European Space Agency (Sentinel-1, Copernicus), NASA/USGS (Landsat, SRTM), ECMWF (ERA5-Land), UCSB (CHIRPS), and NOAA (STORM Events Database). Quantum circuit simulations were performed using PennyLane~\cite{bergholm2018pennylane}. P.N.M.U.H.\ acknowledges support from Lincoln University College, Malaysia. This research received no external funding.
\end{acknowledgments}



\begin{thebibliography}{38}

\bibitem{tellman2021satellite}
B.~Tellman, J.~A.~Sullivan, C.~Kuhn, A.~J.~Kettner, C.~S.~Doyle, G.~R.~Brakenridge, T.~A.~Erickson, and D.~A.~Slayback,
``Satellite imaging reveals increased proportion of population exposed to floods,''
\textit{Nature} \textbf{596}, 80--86 (2021).

\bibitem{toro2001shock}
E.~F.~Toro,
\textit{Shock-Capturing Methods for Free-Surface Shallow Flows}
(John Wiley \& Sons, Chichester, 2001).

\bibitem{mosavi2018flood}
A.~Mosavi, P.~Ozturk, and K.~W.~Chau,
``Flood prediction using machine learning models: Literature review,''
\textit{Water} \textbf{10}, 1536 (2018).

\bibitem{raissi2019physics}
M.~Raissi, P.~Perdikaris, and G.~E.~Karniadakis,
``Physics-informed neural networks: A deep learning framework for solving forward and inverse problems involving nonlinear partial differential equations,''
\textit{J. Comput. Phys.} \textbf{378}, 686--707 (2019).

\bibitem{krishnapriyan2021characterizing}
A.~S.~Krishnapriyan, A.~Gholami, S.~Zhe, R.~M.~Kirby, and M.~W.~Mahoney,
``Characterizing possible failure modes in physics-informed neural networks,''
in \textit{Advances in Neural Information Processing Systems (NeurIPS)} (2021), Vol.~34.

\bibitem{cerezo2021variational}
M.~Cerezo, A.~Arrasmith, R.~Babbush, S.~C.~Benjamin, S.~Endo, K.~Fujii, J.~R.~McClean, K.~Mitarai, X.~Yuan, L.~Cincio, and P.~J.~Coles,
``Variational quantum algorithms,''
\textit{Nat. Rev. Phys.} \textbf{3}, 625--644 (2021).

\bibitem{benedetti2019parameterized}
M.~Benedetti, E.~Lloyd, S.~Sack, and M.~Fiorentini,
``Parameterized quantum circuits as machine learning models,''
\textit{Quantum Sci. Technol.} \textbf{4}, 043001 (2019).

\bibitem{havlicek2019supervised}
V.~Havl\'{\i}\v{c}ek, A.~D.~C\'{o}rcoles, K.~Temme, A.~W.~Harrow, A.~Kandala, J.~M.~Chow, and J.~M.~Gambetta,
``Supervised learning with quantum-enhanced feature spaces,''
\textit{Nature} \textbf{567}, 209--212 (2019).

\bibitem{schuld2021machine}
M.~Schuld and F.~Petruccione,
\textit{Machine Learning with Quantum Computers}, 2nd~ed.
(Springer, Cham, 2021).

\bibitem{klement2026explaining}
N.~Klement, V.~Eyring, and M.~Schwabe,
``Explaining the advantage of quantum-enhanced physics-informed neural networks,''
arXiv:2601.15046 (2026).

\bibitem{dutta2024aqpinns}
S.~Dutta, N.~Innan, S.~Ben~Yahia, and M.~Shafique,
``AQ-PINNs: Attention-enhanced quantum physics-informed neural networks for carbon-efficient climate modeling,''
arXiv:2409.01626 (2024).

\bibitem{grzesiak2024flood}
M.~Grzesiak and P.~Thakkar,
``Flood prediction using classical and quantum machine learning models,''
arXiv:2407.01001 (2024).

\bibitem{farea2025qcpinn}
A.~Farea, S.~Khan, and M.~S.~Celebi,
``QCPINN: Quantum-classical physics-informed neural networks for solving PDEs,''
arXiv:2503.16678 (2025).

\bibitem{kyriienko2021solving}
O.~Kyriienko, A.~E.~Paine, and V.~E.~Elfving,
``Solving nonlinear differential equations with differentiable quantum circuits,''
\textit{Phys. Rev. A} \textbf{103}, 052416 (2021).

\bibitem{kyriienko2023gridfree}
O.~Kyriienko and V.~E.~Elfving,
``Protocols for solving nonlinear ODEs without grid evaluation,''
arXiv:2308.01827 (2023).

\bibitem{kyriienko2024probabilistic}
O.~Kyriienko \textit{et al.},
``Probabilistic quantum physics-informed neural networks,''
\textit{Phys. Rev. Research} \textbf{6}, 033291 (2024).

\bibitem{stein2024hybrid}
S.~A.~Stein \textit{et al.},
``Hybrid quantum physics-informed neural network: Towards efficient learning of high-speed flows,''
arXiv:2503.02202 (2025).

\bibitem{mari2020transfer}
A.~Mari, T.~R.~Bromley, J.~Izaac, M.~Schuld, and N.~Killoran,
``Transfer learning in hybrid classical-quantum neural networks,''
\textit{Quantum} \textbf{4}, 340 (2020).

\bibitem{rodriguez2025satellite}
P.~Rodriguez-Grasa, R.~Farzan-Rodriguez, G.~Novelli, Y.~Ban, and M.~Sanz,
``Satellite image classification with neural quantum kernels,''
\textit{Mach. Learn.: Sci. Technol.} \textbf{6}, 015043 (2025).

\bibitem{mitarai2018quantum}
K.~Mitarai, M.~Negoro, M.~Kitagawa, and K.~Fujii,
``Quantum circuit learning,''
\textit{Phys. Rev. A} \textbf{98}, 032309 (2018).

\bibitem{schuld2019evaluating}
M.~Schuld, V.~Bergholm, C.~Gogolin, J.~Izaac, and N.~Killoran,
``Evaluating analytic gradients on quantum hardware,''
\textit{Phys. Rev. A} \textbf{99}, 032331 (2019).

\bibitem{bergholm2018pennylane}
V.~Bergholm \textit{et al.},
``PennyLane: Automatic differentiation of hybrid quantum-classical computations,''
arXiv:1811.04968 (2018).

\bibitem{lin2017focal}
T.~Y.~Lin, P.~Goyal, R.~Girshick, K.~He, and P.~Doll\'{a}r,
``Focal loss for dense object detection,''
in \textit{Proceedings of the IEEE International Conference on Computer Vision (ICCV)} (2017), pp.\ 2980--2988.

\bibitem{chow1959open}
V.~T.~Chow,
\textit{Open-Channel Hydraulics}
(McGraw-Hill, New York, 1959).

\bibitem{gal2016uncertainty}
Y.~Gal and Z.~Ghahramani,
``Dropout as a Bayesian approximation: Representing model uncertainty in deep learning,''
in \textit{Proceedings of the 33rd International Conference on Machine Learning (ICML)} (2016), pp.\ 1050--1059.

\bibitem{mcclean2018barren}
J.~R.~McClean, S.~Boixo, V.~N.~Smelyanskiy, R.~Babbush, and H.~Neven,
``Barren plateaus in quantum neural network training landscapes,''
\textit{Nat. Commun.} \textbf{9}, 4812 (2018).

\bibitem{cerezo2021cost}
M.~Cerezo, A.~Sone, T.~Volkoff, L.~Cincio, and P.~J.~Coles,
``Cost function dependent barren plateaus in shallow parametrized quantum circuits,''
\textit{Nat. Commun.} \textbf{12}, 1791 (2021).

\bibitem{hewage2026hybrid}
P.~N.~M.~Ukwatta~Hewage, M.~Chakkravarthy, and R.~K.~Abeysekara,
``A hybrid AI framework for multi-modal flood prediction: Integrating Bayesian neural networks, physics-informed constraints, and multi-task learning,''
\textit{Nat. Hazards} (2026, submitted).

\bibitem{kingma2015adam}
D.~P.~Kingma and J.~Ba,
``Adam: A method for stochastic optimization,''
in \textit{Proceedings of the 3rd International Conference on Learning Representations (ICLR)} (2015).

\bibitem{kim2023evidence}
Y.~Kim \textit{et al.},
``Evidence for the utility of quantum computing before fault tolerance,''
\textit{Nature} \textbf{618}, 500--505 (2023).

\bibitem{temme2017error}
K.~Temme, S.~Bravyi, and J.~M.~Gambetta,
``Error mitigation for short-depth quantum circuits,''
\textit{Phys. Rev. Lett.} \textbf{119}, 180509 (2017).

\bibitem{farhi2018classification}
E.~Farhi and H.~Neven,
``Classification with quantum neural networks on near term processors,''
arXiv:1802.06002 (2018).

\bibitem{peruzzo2014variational}
A.~Peruzzo, J.~McClean, P.~Shadbolt, M.-H.~Yung, X.-Q.~Zhou, P.~J.~Love, A.~Aspuru-Guzik, and J.~L.~O'Brien,
``A variational eigenvalue solver on a photonic quantum processor,''
\textit{Nat. Commun.} \textbf{5}, 4213 (2014).

\bibitem{liu2023physicsinformed}
Y.~Liu \textit{et al.},
``Physics-informed graph neural networks for flood forecasting,''
\textit{J. Hydrol.} \textbf{620}, 129376 (2023).

\bibitem{senanayake2022flood}
S.~Senanayake, B.~Pradhan, A.~Alamri, and H.~J.~Park,
``A new application of deep neural network (LSTM) and RUSLE for soil erosion prediction,''
\textit{Geocarto Int.} \textbf{37}, 1--23 (2022).

\bibitem{twele2016sentinel}
A.~Twele, W.~Cao, S.~Plank, and S.~Martinis,
``Sentinel-1-based flood mapping: A fully automated processing chain,''
\textit{Int. J. Remote Sens.} \textbf{37}, 2990--3004 (2016).

\bibitem{otgonbaatar2022classification}
S.~Otgonbaatar and M.~Datcu,
``Classification of remote sensing images with parameterized quantum gates,''
\textit{IEEE Geosci. Remote Sens. Lett.} \textbf{19}, 1--5 (2022).

\bibitem{delilbasic2021quantum}
A.~Delilbasic, G.~Cavallaro, M.~Willsch, F.~Melgani, M.~Riedel, and K.~Michielsen,
``Quantum support vector machine algorithms for remote sensing data classification,''
in \textit{2021 IEEE International Geoscience and Remote Sensing Symposium (IGARSS)} (2021), pp.\ 2608--2611.

\bibitem{abbas2021power}
A.~Abbas, D.~Sutter, C.~Zoufal, A.~Lucchi, A.~Figalli, and S.~Woerner,
``The power of quantum neural networks,''
\textit{Nat. Comput. Sci.} \textbf{1}, 403--409 (2021).

\bibitem{qpinn2024entropy}
I.~Garc\'{i}a-Barrenechea, S.~Borr\`{a}s, and J.~Latorre,
``Quantum physics-informed neural networks,''
\textit{Entropy} \textbf{26}, 649 (2024).

\bibitem{tezuka2025trainable}
H.~Tezuka \textit{et al.},
``Trainable embedding quantum physics informed neural networks,''
\textit{Sci. Rep.} \textbf{15}, 3894 (2025).

\end{thebibliography}
\end{document}